\documentclass[aps,prb,reprint,superscriptaddress]{revtex4-2}
\usepackage{graphicx}
\begin{document}

\title{Raman and infrared studies of CdSe/CdS core/shell nanoplatelets}


\author{Alexander~I.~Lebedev}
\affiliation{Physics Department, Moscow State University, 119991 Moscow, Russia}
\email[]{swan@scon155.phys.msu.ru}
%\thanks{}
\author{Bedil~M.~Saidzhonov}
\affiliation{Department of Materials Science, Moscow State University, 119991 Moscow, Russia}
\author{Konstantin~A.~Drozdov}
\affiliation{Physics Department, Moscow State University, 119991 Moscow, Russia}
\author{Andrey~A.~Khomich}
\affiliation{Institute of Radio Engineering and Electronics RAS, 141190 Fryazino, Moscow region, Russia}
\author{Roman~B.~Vasiliev}
\affiliation{Department of Materials Science, Moscow State University, 119991 Moscow, Russia}
\affiliation{Chemistry Department, Moscow State University, 119991 Moscow, Russia}

\date{\today}

\begin{abstract}
The vibrational spectroscopy of semiconductor nanostructures can provide important
information on their structure. In this work, experimental Raman and infrared
spectra are compared with vibrational spectra of CdSe/CdS core/shell nanoplatelets
calculated from first principles using the density functional theory. The
calculations confirm the two-mode behavior of phonon spectra of nanostructures.
An analysis of the experimental spectra reveals the absence of modes with a high
amplitude of vibrations of surface atoms, which indicates their strong damping.
Taking into account the difference in the damping of different modes and their
calculated intensities, all bands in the spectra are unambiguously identified.
It is found that the frequencies of longitudinal optical modes in heterostructures
are close to the frequencies of LO phonons in bulk strained constituents, whereas
the frequencies of transverse modes can differ significantly from those of the
corresponding TO phonons. It is shown that an anomalous thickness dependence of
CdS TO mode is due to a noticeable surface relaxation of the outer Cd layer in
the nanostructure.

\texttt{J.~Phys.~Chem.~C 125, 6758 (2021); DOI: 10.1021/acs.jpcc.0c10529}

\end{abstract}

%\keywords{}

\maketitle

\section{Introduction}

In recent years, colloidal quasi-two-dimensional (2D) nanoplatelets of cadmium
chalcogenides and, in particular, cadmium selenide CdSe, have become the object
of intense investigations due to the unique optical and electronic properties
of these materials~\cite{ChemRev.116.10934}. Because of the quantization of
carriers in one direction and atomically smooth surfaces, these low-dimensional
structures are characterized by extremely narrow luminescence
bands~\cite{NatureMater.10.936}, a giant oscillator strength of optical
transitions~\cite{PhysRevB.91.121302}, and associated with it large absorption
cross sections~\cite{JPhysChemC.119.26768} and fast recombination dynamics,
which makes them promising materials for LEDs~\cite{Materials.11.1376},
photodetectors~\cite{NanoLett.15.1736}, and lasers~\cite{JPhysChemC.122.10659}.
However, the photoluminescence quantum yield of nanoplatelets made of a single
semiconductor is relatively small, and their embedding into various media
leads to further degradation of their optical properties, which limits their
applications.

To increase the quantum yield and stability of the samples, it was proposed to
use heterostructures with a shell made of a wide-gap semiconductor. The methods
of synthesis of the core/shell~\cite{JAmChemSoc.134.18591},
core/crown~\cite{JAmChemSoc.135.14476, NanoLett.14.207}, and core/crown/shell
structures~\cite{AdvFunctMater.26.3570} were then developed. The photoluminescence
quantum yield of such heterostructures can reach 98\%~\cite{Small.15.1804854}.
By varying the composition and thickness of the cladding material, it is possible
to tune the position of absorption and luminescence bands of the nanostructures
in a wide spectral range~\cite{JLumin.209.170}. In contrast to quantum dots,
nanoplatelets and nanoheterostructures obtained by the layer-by-layer deposition
on the surfaces of nanoplatelets are homogeneous in thickness, and so are very
attractive both for fundamental studies and different applications because of
the absence of inhomogeneous broadening of their spectra. One of such widely
studied systems are nanostructures with the CdSe/CdS heterojunction
\cite{NanoLett.3.1677, JAmChemSoc.125.12567, NanoLett.6.463, Nanotechnol.18.285701,
JAmChemSoc.135.14476, JPhysChemC.117.18225, SuperlattMicrostr.98.158,
NatureCommun.8.143, OptMater.82.135, Nanotechnology.29.395604, JPhysChemLett.9.286}.

The electronic and optical properties of semiconductor nanoheterostructures depend
not only on their composition, size, and morphology, but also on the structural
stresses arising from the lattice mismatch of constituent materials. In addition,
in real quasi-2D structures, surface structural relaxations and the effect
of surface ligands on mechanical stresses in nanostructures can appear. Therefore,
when developing materials based on quasi-2D heterostructures, it is necessary to
have complete information about their real structure. In this regard, along with
traditional optical spectroscopy, vibrational spectroscopy of semiconductor
nanostructures~\cite{JPhysD.51.503001} can provide valuable information on their
structure, composition~\cite{JPhysCondensMatter.16.9069},
sizes~\cite{NanoLett.8.4614}, mechanical stresses~\cite{JPhysChemC.117.18225},
and on the geometry of the surface layer. Despite a large number of publications
devoted to quasi-2D structures, systematic studies of the lattice dynamics of both
single-phase nanoplatelets and heterostructures based on them has been limited
by few works~\cite{PhysRevB.88.041303, PhysRevLett.113.087402, PhysRevB.96.184306,
JPhysChemC.122.27100, Nanoscale.8.17204, Langmuir.34.13828}. In
Ref.~\cite{Nanoscale.8.17204}, comparative analysis of vibrational spectra of
CdSe/CdS nanostructures of the core/shell and core/crown types was carried out.
Recently, it was proposed that the ZnS shell formation and its thickness in
quasi-2D CdSeS/ZnS core/shell nanostructures can be monitored by Raman
spectroscopy~\cite{Langmuir.34.13828}.

In this work, the vibrational spectra of the CdSe/CdS core/shell nanoplatelets,
which are experimentally studied using Raman and infrared (IR) spectroscopy, are
compared with results of first-principles calculations of the lattice dynamics of
these nanoheterostructures. The absence of many calculated modes in the
experimental spectra is found to correlate with a high amplitude of vibrations
of surface atoms for these modes, which indicates their strong damping. Taking
into account this difference in the damping, all bands in the spectra are
unambiguously identified. It is shown that the observed frequencies are in good
agreement with the frequencies calculated for heterostructures in which the
effects of internal strains and the relaxation of the surface layers are taken
into account.

%\section{Experimental and computational details}
\section{Methods}

\begin{figure}
\centering
\includegraphics[scale=1.0]{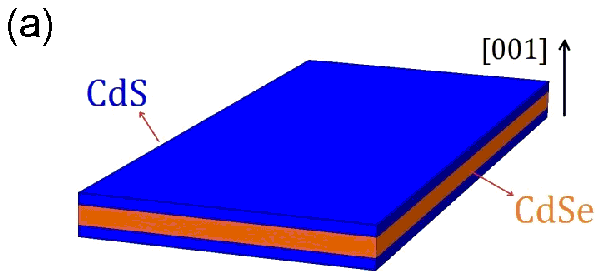}
\centering
\includegraphics[scale=1.0]{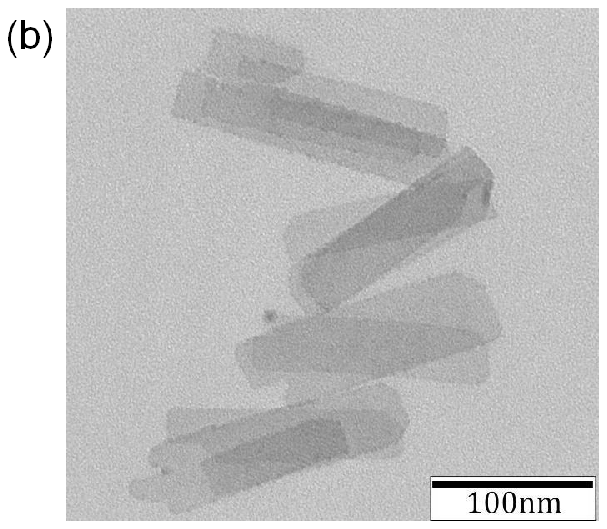}
\caption{\label{fig1}(a) Geometry of CdSe/CdS core/shell nanoplatelets and
(b) their TEM image.}
\end{figure}

The CdSe/CdS core/shell nanoplatelets studied in this work were ultrathin
nanoplatelets whose geometry is shown in Fig.~\ref{fig1}(a). The middle part of
these nanoheterostructures---quasi-two-dimensional CdSe nanoplatelets---served
as a base for the following growth of CdS shell and were synthesized by colloidal
method~\cite{ChemMater.29.579}. The surface of the nanoplatelets obtained in
this way is terminated by cadmium atoms. A~CdS shell with a thickness from one
to three monolayers (ML) was grown on the surfaces of these nanoplatelets by the
layer-by-layer deposition in a polar solvent (N-methylformamide,
NMF)~\cite{JLumin.209.170}. To grow a sulfur monolayer, 1~ml of 0.1~M sodium
sulfide solution in NMF was added to 1~ml of nanocrystal solution in hexane.
Replacing the stabilizer of CdSe nanoplatelets (oleic acid) with a S monolayer
results in the transition of nanoplatelets from a non-polar to a polar solvent.
After complete transfer of nanoplatelets into NMF, the nanoplatelets were
precipitated by adding toluene and centrifuging at 6000~rpm, then washed several
times with acetone. Next, 1~ml of a 0.2~M solution of cadmium acetate in
NMF was added to the resulting precipitate and left on for 40~min to form a
cationic monolayer, after which the nanoplatelets were precipitated with a mixture
of toluene and acetonitrile (1:1 by volume), separated by centrifugation, and
dissolved in pure NMF. The above procedure describes the formation of one CdS
monolayer on the surface of CdSe nanoplatelets. To prepare CdSe/2CdS and CdSe/3CdS
core/shell nanoplatelets, this procedure was repeated two and three times,
respectively. The surface of the obtained nanostructures was then stabilized
with oleic acid, for which they were dissolved in a mixture of toluene and oleic
acid. After that, the nanoplatelets were precipitated with acetone and dissolved
in a suitable solvent.

The obtained samples were characterized by X-ray diffraction, optical
absorption spectroscopy, and transmission electron microscopy (TEM). X-ray
studies were performed on D/Max 2500V/PC Rigaku diffractometer using
Cu $K_{\alpha}$ radiation. The LEO912 AB OMEGA system with an accelerating
voltage of 100 keV was used to obtain TEM images. The samples for these studies
were prepared by evaporation of hexane dispersion of nanoplatelets on
monocrystalline (001) Si substrates or on the copper electron microscopy grids,
respectively.

The vibrational spectra of the samples were studied by Raman and IR
spectroscopies at room temperature. The samples for these studies were prepared
by depositing the nanocrystal solution on the surface of silicon substrates.
IR transmission spectra were recorded using a VERTEX~70v Fourier spectrometer
in the frequency range 50--700~cm$^{-1}$. Raman spectra were recorded on a Jobin
Yvon LabRAM HR800 spectrometer with excitation by a semiconductor laser with
$\lambda = 473$~nm. Strong Raman signals were observed only when the excitation
energy exceeded the energy of the absorption edge of the nanostructures.

To identify bands in the vibrational spectra of the core/shell nanoplatelets,
the equilibrium geometry, frequencies of normal modes, high-frequency dielectric
constant, oscillator strengths, and Raman polarizability tensors necessary
for modeling Raman and IR spectra were calculated from first principles.
The calculations were performed using the \texttt{ABINIT} software package,
the plane-wave basis, and norm-conserving pseudopotentials generated by
A.~Khein and D.~C.~Allen using the Trouiller--Martins scheme in the LDA
approximation and taken from~\cite{abinitweb}. The cutoff energy was 50~Ha
(1360~eV), integration over the Brillouin zone was carried out using
a 8$\times$8$\times$2 Monkhorst--Pack mesh. The relaxation of the lattice
parameters and atomic positions was carried out until the Hellmann--Feynman forces
acting on the atoms became less than $5 \cdot 10^{-6}$~Ha/Bohr (0.25~meV/{\AA}).
As in Ref.~\cite{PhysRevB.96.184306}, the electrical activity of the excess cadmium
layer was compensated by surface F~atoms occupying the positions of the virtual
chalcogen atoms in the zinc-blende structure. The modeling of
nanoplatelets was performed on supercells, to which a 20~{\AA} vacuum gap was added
to isolate the nanoplatelets from each other. The symmetry of such supercells is
described by the space group $P{\bar 4}m2$.%
    \footnote{Strictly speaking, the actual symmetry of a single nanoplatelet
    is described by the $p$--$4m2$ layer group. The symmetries and irreducible
    representations for this group can be obtained using the \texttt{LSITESYM}
    program~\cite{JApplCryst.52.1214} on the Bilbao Crystallographic Server
    from the positions of atoms. Our tests showed that both approaches, using
    the $P{\bar 4}m2$ space group and the $p$--$4m2$ layer group, give the same
    results for the symmetry of vibrations in our nanoplatelets.}
Calculations of theoretical Raman and IR spectra were carried out assuming
a random spatial orientation of nanoplatelets.

\section{Results}

\subsection{Experimental results}

According to TEM data, the CdSe/CdS core/shell nanoplatelets have a perfect
flat rectangular shape with lateral dimensions of about 150$\times$40~nm and a
thickness of 1.5--3.3~nm (Fig.~\ref{fig1}(b)). As shown earlier~\cite{JLumin.209.170},
the nanoplatelets have a zinc-blende structure (see Fig.~S1 in the Supporting
Information) and are oriented normal to the [001]~axis (Fig.~\ref{fig1}(a)).
Although the basic CdSe nanoplatelets underwent spontaneous
folding~\cite{ChemMater.29.579} (see Fig.~S2 in the Supporting Information),
all CdSe/CdS core/shell nanostructures have a flat shape independent of the shell
thickness.

\begin{figure}[t]
\centering
\includegraphics{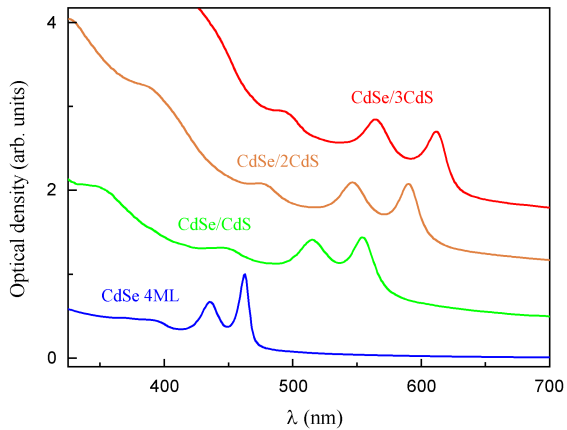}
\caption{\label{fig2}Absorption spectra of quasi-2D CdSe nanoplatelets and CdSe/CdS
core/shell nanostructures.}
\end{figure}

The absorption spectra of CdSe nanoplatelets and CdSe/CdS core/shell nanostructures
are shown in Fig.~\ref{fig2}. The basic CdSe nanoplatelets have the main exciton
absorption peak at 463~nm. With an increase in the shell thickness, a red shift
of the absorption bands is observed, which, according to
Ref.~\cite{Nanoscale.8.17204}, indicates that the CdSe/CdS heterostructures
are of the core/shell type. Taking into account that in nanoplatelets with an
absorption peak at 463~nm, the energy of the lowest-frequency vibrational mode
in the Raman spectra was 40~cm$^{-1}$~\cite{Nanoscale.8.13251}, from the
calculations of Ref.~\cite{PhysRevB.96.184306} we conclude that the band at this
frequency corresponds to the nanoplatelet thickness of 4~ML (4.5~ML if one adds
the thickness of an additional Cd layer on the surface). This was a reason why
the following calculations of vibrational spectra of nanoheterostructures were
performed for this thickness of the CdSe layer.

\begin{figure}
\centering
\includegraphics{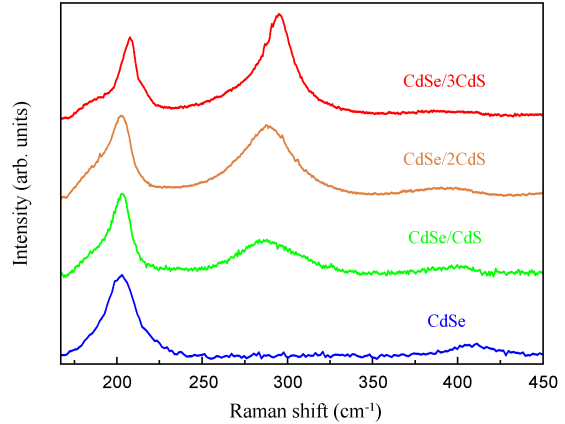}
\caption{\label{fig3}Raman spectra of quasi-2D CdSe nanoplatelets and CdSe/CdS
core/shell nanostructures.}
\end{figure}

\begin{figure}
\centering
\includegraphics{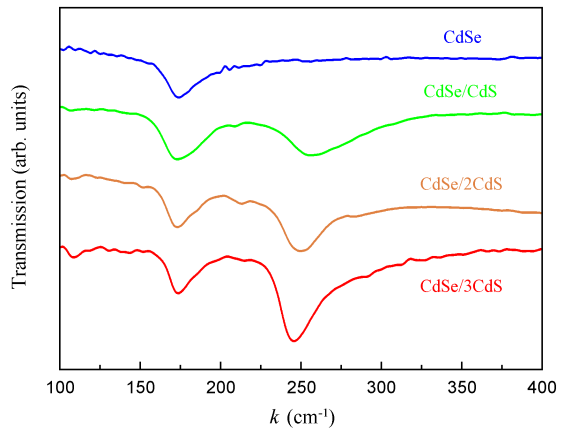}
\caption{\label{fig4}Infrared spectra of quasi-2D CdSe nanoplatelets and CdSe/CdS
core/shell nanostructures.}
\end{figure}

Experimental Raman and IR spectra of CdSe/CdS core/shell nanoplatelets with different
shell thickness are shown in Figs.~\ref{fig3} and \ref{fig4}. The Raman spectrum
of the basic CdSe nanoplatelets contains a weakly asymmetric band at about
200~cm$^{-1}$ and a weak band at about 400~cm$^{-1}$. In Ref.~\cite{PhysRevB.88.041303},
the band at 200~cm$^{-1}$ was interpreted as two overlapping peaks corresponding
to the longitudinal
optical (LO) mode and the surface optical (SO) mode, while the band at 400~cm$^{-1}$
was associated with the second-order Raman scattering. In Ref.~\cite{PhysRevB.96.184306},
another interpretation based on calculations of the vibrational spectra of CdSe
nanoplatelets was proposed. In this interpretation, the band at 200~cm$^{-1}$ is
associated with a vibrational mode of the $A_1$~symmetry, which makes a stronger
contribution into the Raman spectrum as compared to the LO mode of the
$B_2$~symmetry, and the mode previously attributed to the surface SO mode is merely
a different mode of the $A_1$ symmetry (recall that in classical electrodynamics
the frequencies of SO~modes in thin films with a flat surface coincide with the
frequencies of bulk material~\cite{RepProgrPhys.33.149}).

The addition of CdS shell results in the appearance of a new strong band in the
Raman spectra at about 280~cm$^{-1}$, whose intensity and frequency increase with
increasing shell thickness. This partially agrees with the results of the study
of CdSe/CdS core/shell nanostructures~\cite{Nanoscale.8.17204}, where a band with
a complex structure was observed in the 270--300~cm$^{-1}$ region. This band
shifted toward higher frequencies with increasing shell thickness and was attributed
to the LO phonon in CdS and interface modes arising at the interface between
two materials.

The infrared spectrum of the CdSe nanoplatelet is characterized by one asymmetric
band at about 175~cm$^{-1}$ (Fig.~\ref{fig4}). After adding the CdS shell, a new
asymmetric band, whose intensity increases and whose energy decreases with
increasing shell thickness, appears in the IR spectra at about 250~cm$^{-1}$.
In addition,
the IR spectra exhibit a weak feature at about 210~cm$^{-1}$, whose frequency
weakly increases with increasing shell thickness, and a very weak feature around
290~cm$^{-1}$. The obtained spectra are consistent with the IR reflection spectrum
reported in Ref.~\cite{Nanoscale.8.17204}, in which the absorption bands were
interpreted as manifestations of TO, SO, and LO phonons of two components of
the nanoheterostructure.%
    \footnote{We note that the IR spectrum presented in Ref.~\cite{Nanoscale.8.17204}
    is more typical of transmission spectra, rather than of the reflection ones.}

\subsection{Vibrational spectra of CdSe/CdS core/shell nanostructures}

\begin{figure}
\centering
\includegraphics{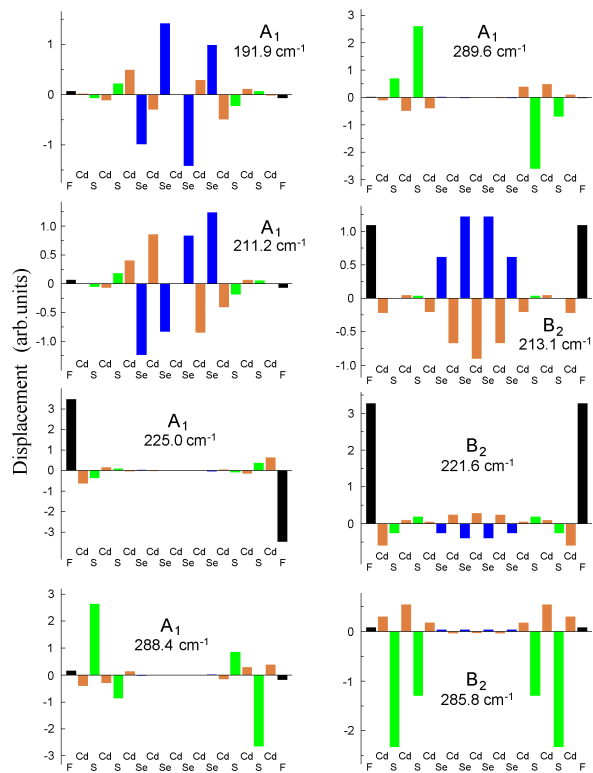}
\caption{\label{fig5}Eigenvectors of all strong Raman active modes of $A_1$ and $B_2$
symmetry for a free-standing CdSe/2CdS nanoheterostructure. Atomic displacements
are normal to the plane of the nanoplatelet. Colored bars denote: brown~---
Cd atoms, blue~--- Se atoms, green~--- S atoms, and black~--- F atoms.}
\end{figure}

\begin{figure}
\centering
\includegraphics{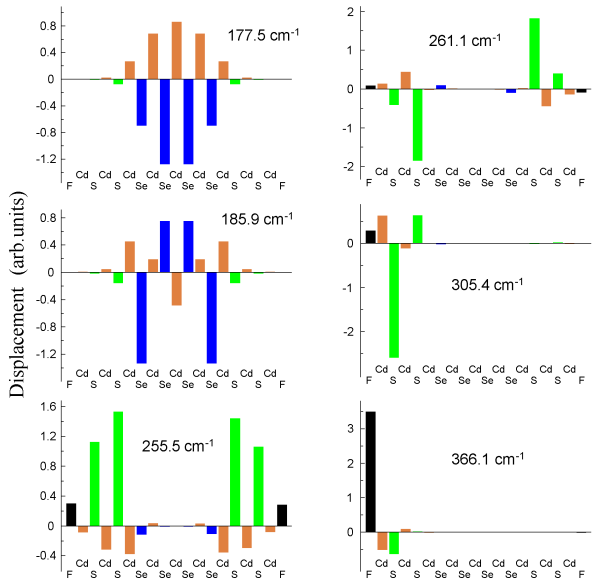}
\caption{\label{fig6}Eigenvectors of all strong IR-active modes of $E$ symmetry for
a free-standing CdSe/2CdS nanoheterostructure. Atomic displacements lie in the
plane of the nanoplatelet. Colored bars denote: brown~--- Cd atoms, blue~---
Se atoms, green~--- S atoms, and black~--- F atoms.}
\end{figure}

First-principles calculations of vibrational spectra of CdSe/CdS core/shell
nanostructures show that the symmetries of vibrational modes in them are similar
to those in previously studied CdSe nanoplatelets~\cite{PhysRevB.96.184306}. They
are $A_1$ and $B_2$ for quasi-Lamb modes with atomic displacements normal to the
nanoplatelet, and $E$ for in-plane atomic displacements. In the high-frequency
region (optic-like modes, $>$170~cm$^{-1}$) in the basic CdSe nanoplatelet
with a thickness of 4ML there are three $A_1$~modes, three $B_2$~modes, and five
$E$~modes~\cite{ PhysRevB.96.184306}. With the addition of each extra CdS monolayer
on both sides of the basic nanoplatelet, the number of optic-like modes increases
by one $A_1$~mode, one $B_2$~mode, and two $E$~modes. Although all three modes
are Raman active, the main contribution to the Raman spectra, like in CdSe
nanoplatelets~\cite{PhysRevB.96.184306}, comes from the $A_1$~modes; the contribution
of the $B_2$~modes is small, and the contribution of the $E$~modes is extremely
small. In infrared spectra, the $B_2$ and $E$~modes are active in the out-of-plane
and in-plane polarization, respectively.

The eigenvectors of $A_1$ and $B_2$~modes in the CdSe/2CdS nanoheterostructure
are shown in Fig.~\ref{fig5}, and the eigenvectors of the $E$~mode are shown
in Fig.~\ref{fig6}. An analysis of these figures shows that after excluding
the surface modes with frequencies 221.6, 225.0, 305.4, and 366.1~cm$^{-1}$,
the figures clearly demonstrate two types of vibrations, in which the atomic
displacements are confined either in the CdSe layer or in the CdS one. This
agrees with the results of an experimental study of phonon spectra of CdSe--CdS
solid solutions~\cite{PhysRev.155.750} and the results of our calculations
of phonon spectra of superlattices and disordered solid solutions in the
CdSe--CdS system (see Figs.~S3--S6 in the Supporting Information), which
all indicate its two-mode behavior. Physically, this
means that vibrations with a given frequency can propagate only in one of
the two components of the heterostructure and, therefore, they are localized
in the corresponding region of the nanoplatelet. No interface modes discussed
in Ref.~\cite{Nanoscale.8.17204} were found when analyzing the eigenvectors of
phonons in CdSe/CdS core/shell nanostructures and CdSe/CdS superlattices (see
Fig.~S5 and S6 in the Supporting Information).

The displacement patterns of the $E$~mode with a frequency of 177.5~cm$^{-1}$
and the $B_2$~mode with a frequency of 213.1~cm$^{-1}$ are closest to those of
optical phonons in bulk CdSe. The frequency of the first of these modes is
close to the calculated frequency of TO phonon in the bulk material
(174.8~cm$^{-1}$), whereas the second frequency differs markedly from that of
LO phonon (204.5~cm$^{-1}$). The displacement patterns of the $E$~mode with a
frequency of 255.5~cm$^{-1}$ and the $B_2$~mode with a frequency of 285.8~cm$^{-1}$
are similar to those of TO and LO modes in CdS, but their frequencies are strongly
different from the frequencies of the corresponding phonons in bulk CdS (243.9
and 296.4~cm$^{-1}$). However, it should be taken into account that both
layers in the heterostructure are strained as a result of a mismatch in the
lattice parameters of CdSe and CdS.

The calculation of the frequencies of phonons in biaxially strained bulk CdS
and CdSe, whose in-plane lattice parameter is equal to that of the core/shell
nanostructure, gave the following frequencies of these modes: 176.0 (TO$_{xy}$)
and 216.6 (LO$_z$) for CdSe and 243.4 (TO$_ {xy}$) and 290.1~cm$^{-1}$ (LO$_z$)
for CdS. It is seen that a proper account of strain effects improves the
agreement between the frequencies of the LO$_z$ modes in CdS and CdSe and the
TO$_{xy}$ mode in CdSe with the frequencies calculated for the
nanoheterostructure. However, the agreement for the TO$_{xy}$ mode in CdS
is not good. This means that, in addition to strain, there is another effect
that strongly influences the vibrational spectra of nanoheterostructures.

The frequency of the aforementioned ``anomalous'' mode, which was attributed
to the TO$_{xy}$ vibrations in the CdS layer, approaches the TO phonon frequency
of bulk CdS when increasing shell thickness: it is 266.1~cm$^{-1}$ in the
core/shell nanostructure with one CdS monolayer, 255.5~cm$^{-1}$ in the structure
with two monolayers, and 250.9+252.9~cm$^{-1}$ (two modes differing in the
atomic displacement pattern) in the structure with three monolayers. An analysis
shows that the observed effect is due to a noticeable relaxation of the
surface Cd layer in the structure. As follows from the atomic positions given
in Tables~S1--S3 in the Supporting Information, the Cd(2)--S(3) distance is
shortened by $\sim$1.5\% (0.036--0.041~{\AA}) as compared to other Cd--S distances
in the structure. The Gr{\"u}neisen parameter $(1/\omega)d\omega/da$ for this
confined mode is quite large ($\sim$$5.6$), and so its frequency in
one-monolayer-thick CdS layer can be increased easily by 20~cm$^{-1}$, in
agreement with our predictions and experiment. The change in frequency of
collective vibrations of thicker CdS layers is roughly proportial to the inverse
of the layer thickness.

\begin{figure}
\centering
\includegraphics{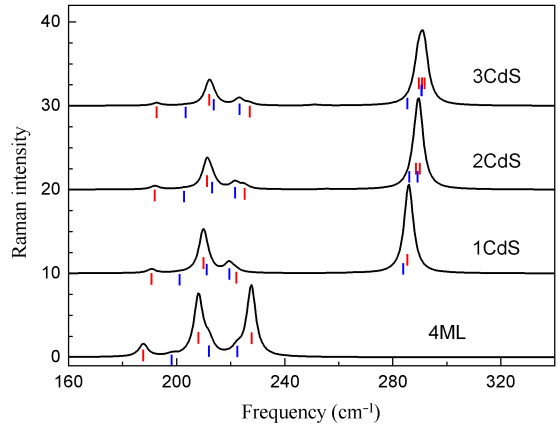}
\caption{\label{fig7}Raman spectra of free-standing CdSe/CdS nanoheterostructures
with different thickness of the CdS layer calculated under the assumption of the
same damping for all modes. Red vertical strokes indicate modes with the
$A_1$~symmetry, blue vertical strokes indicate the $B_2$~modes.}
\end{figure}

\section{Discussion}

\begin{figure}
\centering
\includegraphics{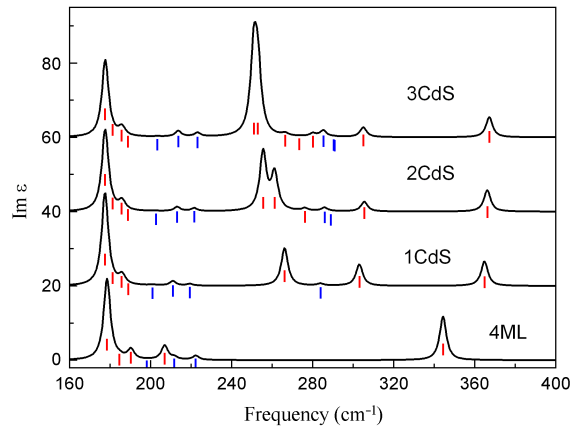}
\caption{\label{fig8}Calculated IR absorption spectra of free-standing CdSe/CdS
nanoheterostructures calculated under the assumption of the same damping for all
modes. The frequencies of the $E$~modes are shown by red vertical strokes,
those of the $B_2$~modes are shown by blue vertical strokes. The highest energy
modes correspond to surface vibrations of F~atoms.}
\end{figure}

The calculated Raman and IR spectra of free-standing CdSe/CdS core/shell
nanostructures with different thickness of the CdS shell are shown in
Figs.~\ref{fig7} and \ref{fig8}. A comparison of these spectra with the
experimental data (Figs.~\ref{fig3} and \ref{fig4}) shows that in the
experimental IR spectra, the high-frequency modes with calculated frequencies
$>$300~cm$^{-1}$ are absent. In the Raman spectra, the same is true for the
227.6~cm$^{-1}$ mode in the basic CdSe nanoplatelet and for modes in the
220--230~cm$^{-1}$ range for core/shell nanostructures. We suppose that the
deficiency of the bands in the experimental spectra is due to a strong damping
of these modes. Indeed, the F~atom in our calculations mimics the oxygen atom
of the carboxyl group of the oleic acid. The molecules of the oleic acid are
attached to the surface of the nanoplatelet by these groups, but their long
tails are disordered and form on the surface of the nanoplatelet a kind of a
liquid layer, which strongly absorbs the high-frequency vibrations of surface
atoms via the volume viscosity mechanism.
This absorption results in a strong damping of modes having a large amplitude
of vibrations of surface F~atoms. The corresponding resonances become very wide
and the modes are practically invisible in the spectra. This
aspect should be taken into account when interpreting the spectra. Because
of different damping, different bands in the spectra can have different
intensities and halfwidths. The parameter characterizing the damping can be
defined as the relative fraction of the vibrational energy of surface
F~atoms in the total vibrational energy of a given mode,
$W_{\rm F} = (\sum_{\rm F} M_{\rm F} u_{\rm F}^2) / \sum_i M_i u_i^2$,
where $M_i$ is the mass and $u_i$ is the atomic displacement of the $i$th
atom ($i$ = F, Cd, Se, S) and the last sum is taken over all atoms in the unit
cell of the nanoplatelet.

\begin{table}
\caption{\label{table1}Frequencies and symmetries of optic-like Raman active
modes in the vibrational spectrum of the CdSe/2CdS nanoheterostructure,
their calculated intensities, and relative vibrational energies of surface
F~atoms.}
\begin{ruledtabular}
\begin{tabular}{cccc}
Symmetry  & Frequency   & Relative  & $W_{\rm F}$ \\
          & (cm$^{-1}$) & intensity & \\
\hline
$A_1$     & 191.9       & 0.427     & $2.6 \cdot 10^{-4}$ \\
$B_2$     & 202.7       & 0.066     & $2.8 \cdot 10^{-4}$ \\
$A_1$     & 211.2       & 3.419     & $2.7 \cdot 10^{-4}$ \\
$B_2$     & 213.1       & 0.721     & 0.082 \\
$B_2$     & 221.6       & 0.816     & 0.739 \\
$A_1$     & 225.0       & 0.475     & 0.824 \\
$B_2$     & 285.8       & 0.049     & $4.4 \cdot 10^{-4}$ \\
$A_1$     & 288.4       & 2.082     & $1.7 \cdot 10^{-3}$ \\
$B_2$     & 288.9       & 0.021     & $1.4 \cdot 10^{-3}$ \\
$A_1$     & 289.6       & 9.402     & $1.1 \cdot 10^{-5}$ \\
\end{tabular}
\end{ruledtabular}
\end{table}

We consider as an example the CdSe/CdS nanoheterostructure with a shell thickness
of two monolayers (Table~\ref{table1}). Optic-like modes in this structure are
represented by $5A_1 + 5B_2 + 9E$~modes. The calculated Raman spectrum of this
nanostructure contains three intense $A_1$~modes (211.2, 288.4, and 289.6~cm$^{-1}$),
four modes of medium intensity, and three weak modes. Of three strongest modes,
the 288.4~cm$^{-1}$ one is characterized by the largest value of $W_{\rm F}$ and
will most likely be indistinguishable from the stronger 289.6~cm$^{-1}$~line.
Of the medium-intensity modes, only the $A_1$~mode with a frequency of
191.9~cm$^{-1}$ is characterized by a small value of $W_{\rm F}$ and it can be
observed in the experimental spectra as a shoulder on the low-energy side of the
211.2~cm$^{-1}$~line. Other three modes of medium intensity are characterized by
high $W_{\rm F}$ values and most likely will contribute to a structureless
background in the experimental spectrum. Weak lines have too low intensity to
be observed in the spectra.

\begin{table}
\caption{\label{table2}Frequencies and symmetries of optic-like infrared active
modes in the vibrational spectrum of the CdSe/2CdS nanoheterostructure, their
oscillator strengths, and relative vibrational energies of surface F~atoms.}
\begin{ruledtabular}
\begin{tabular}{cccc}
Symmetry  & Frequency   & Oscillator strength      & $W_{\rm F}$ \\
          & (cm$^{-1}$) & (10$^{-4}$ atomic units) & \\
\hline
$E$       & 177.5       & 2.14   & $4.3 \cdot 10^{-7}$ \\
$E$       & 181.5       & 0.00   & $1.3 \cdot 10^{-6}$ \\
$E$       & 185.9       & 0.26   & $1.7 \cdot 10^{-6}$ \\
$E$       & 189.1       & 0.00   & $8.2 \cdot 10^{-7}$ \\
$B_2$     & 202.7       & 0.01   & $2.8 \cdot 10^{-4}$ \\
$B_2$     & 213.1       & 0.27   & 0.082 \\
$B_2$     & 221.6       & 0.20   & 0.739 \\
$E$       & 255.5       & 2.22   & 0.012 \\
$E$       & 261.1       & 1.42   & $8.2 \cdot 10^{-4}$ \\
$E$       & 276.2       & 0.11   & 0.026 \\
$B_2$     & 285.8       & 0.30   & $4.4 \cdot 10^{-4}$ \\
$B_2$     & 288.9       & 0.01   & $1.4 \cdot 10^{-3}$ \\
$E$       & 305.4       & 0.44   & $5.6 \cdot 10^{-3}$ \\
$E$       & 366.1       & 1.16   & 0.847 \\
\end{tabular}
\end{ruledtabular}
\end{table}

Consider now the infrared spectra. The calculated IR spectrum contains two pairs
of $E$~modes with high oscillator strength and frequencies of 177.5, 185.9,
255.5, and 261.1~cm$^{-1}$ (Table~\ref{table2}). These lines form two asymmetric
bands observed in the experimental spectrum. The two highest-frequency $E$~modes,
as was noted above, are absent in the spectra because of their strong damping.
The rest of the $E$~modes are too weak to produce detectable features in the
spectra. As for a weak peak observed in the experimental spectra around
210~cm$^{-1}$, it can be related to the brightest of the $B_2$~modes (213.1~cm$^{-1}$)
despite its a rather moderate value $W_{\rm F}$ (the other $B_2$~mode with a
frequency of 221.6~cm$^{-1}$ is a surface mode, which can be neglected because
of its strong damping). The $B_2$~mode corresponding to the LO phonon in CdS
(285.78~cm$^{-1}$) was very weak in our IR spectra presumably because of its
low calculated intensity. This mode was observed in Ref.~\cite{Nanoscale.8.17204}
in the IR spectra of one of the CdSe/CdS core/shell nanoplatelets whose shell
thickness was not reported. These $B_2$~modes, whose polarization is normal to
the surface of the nanoplatelet, become visible in the IR spectra because of the
random orientation of nanoplatelets in samples used for optical measurements.

A~similar analysis was carried out for CdSe/CdS core/shell nanoplatelets with
other thicknesses of the CdS layer (the results of calculations are given in
Tables~S4--S7 in the Supporting Information). It confirmed that modes with the
strongest contribution to the experimental spectra are characterized by rather
small $W_{\rm F}$ values. This observation shows that an analysis of vibrational
spectra combined with first-principles calculations of the eigenvectors of
vibrational modes and their intensities can be a powerful technique for
unambiguous identification of modes in experimental spectra. In our case, it
allowed to identify five strongest modes among about 60 possible vibrations.

\begin{figure}
\centering
\includegraphics{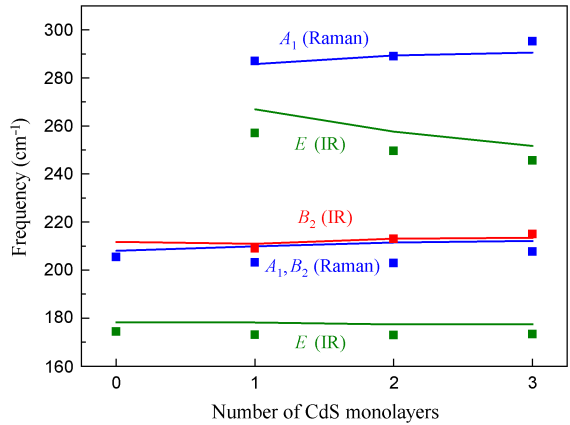}
\caption{\label{fig9}Dependence of vibrational frequencies in CdSe/CdS core/shell
nanoplatelets on the shell thickness. Solid lines show the results of calculations,
points are experimental data.}
\end{figure}

The vibrational frequencies obtained from the experiment are compared with the
results of calculations for all studied nanoheterostructures in Fig.~\ref{fig9}.
For structures with a shell thickness of two and three monolayers, in the figure
we show the center of gravity of two or three $A_1$~modes with close frequencies
and low damping as the frequency of the high-frequency mode (we recall that the
number of these $A_1$~modes coincides with the number of CdS layers).%
    \footnote{As shown in Ref.~\cite{JPhysD.51.503001}, the described fine
    structure in Raman spectra can be resolved only at low temperatures.}
The center of gravity of two $E$~modes with the highest oscillator strengths was
used as the frequency of the high-frequency $E$~mode. According to our
calculations, the frequency of the $B_2$~mode is 1--2~cm$^{-1}$ higher than that
of the $A_1$~mode; this prediction is confirmed by experiment.

An inspection of Fig.~\ref{fig9} finds a slight systematic overestimation of
all calculated frequencies in comparison with experiment. We believe that it may
be a consequence of the LDA approximation used in our calculations because it
slightly underestimates the lattice parameter and overestimates the mode frequencies.
In addition, until now, we have assumed that the nanoplatelets are free-standing.
However, being very thin, they may be strained, for example, under the action
of van der Waals attraction between surface ligands (oleic acid molecules).
Therefore, it is desirable to check whether the structure is strained (we recall
that the internal strains originating from the difference in the lattice parameters
of two materials have already been taken into account).

The in-plane strain of the nanostructure can be estimated from shifts of vibrational
frequencies. Calculations of the frequency shifts $a_0 \cdot (d \omega / da)$ for
the actual modes gave the following results: $-$0.3~cm$^{-1}$/\% for both
$E$~modes, $-$(3.8--4.0)~cm$^{-1}$/\% for the low-frequency $A_1$~mode,
$-$(5.5--5.7)~cm$^{-1}$/\% for the high-frequency $A_1$~mode, and
$-$(5.0--5.5)~cm$^{-1}$/\% for the $B_2$~mode. Comparison of the experimental
and calculated frequencies of the five modes observed in the experiment showed
that the minimum of their root-mean-square deviation is achieved when the
structure is stretched by 0.48--0.66\% from the calculated equilibrium. This value
is comparable with the value of the systematic underestimation of the lattice
parameter in the LDA approximation (according to our calculations, for bulk
CdSe and CdS crystals with a wurtzite structure, the systematic underestimations
were $-$0.69\% and $-$0.57\%, respectively). Thus, according to the vibrational
spectroscopy data, the in-plane strain in our nanoplatelets induced by the oleic
acid surface ligands does not exceed 0.1--0.2\%.

When studying nanoplatelets, many authors observe the phenomenon of their
spontaneous scrolling~\cite{ChemMater.29.579,ChemMater.30.1710,ChemMater.31.9652}.
In this case, the inner and outer layers of the nanoplatelet undergo strain, which
complicates their vibrational spectra. In particular, two CdS layers, which
practically do not interact with each other because of the above-mentioned wave
attenuation in the CdSe layer, should produce two different sets of frequencies.
In scrolled nanoplatelets, the splitting of Raman and IR lines can be observed
if it becomes comparable with the linewidth. At the same time, one can expect
that the position of the center of gravity of these groups of lines will remain
close to the position of the line in an unstrained nanoplatelet, and therefore
the above analysis of the vibrational spectra is valid for scrolled nanoplatelets
too.

\section{Conclusions}

In this work, the vibrational spectra of CdSe/CdS core/shell nanoplatelets with
a CdS shell thickness from one to three monolayers have been calculated from
first principles. The two-mode behavior of their phonon spectrum was established.
It was shown that in the nanoheterostructures the frequencies of longitudinal
modes are close to the frequencies of LO phonons in bulk strained constituents,
while the frequencies of transverse modes can markedly differ from those of
the corresponding TO phonons. A comparison with experiment revealed that
modes with a high amplitude of vibrations of the surface atoms are absent in
the Raman and IR spectra, which indicates their strong damping. Taking into
account the difference in the damping of modes, which was determined by the
relative energy of vibrations of surface atoms, and their intensities, all
bands in the spectra were unambiguously identified. As in CdSe nanoplatelets,
the dominant contribution to the Raman spectra comes from the $A_1$~modes.
The absorption bands in the IR spectra are formed by four $E$~modes and one
$B_2$~mode, for which the optical transition is allowed in a geometry inclined
with respect to the surface of the nanoplatelet. When taking into account the
internal strains resulting from the mismatch in the lattice parameters of two
materials and the relaxation of the surface layers, the calculated spectra
become in good agreement with the experimental spectra of the CdSe/CdS
core/shell nanoplatelets.

\begin{acknowledgments}
The work was financially supported by the Russian Foundation for Basic Research
under Grant 19-03-00481.
\end{acknowledgments}

%\bibliography{all}
\providecommand{\BIBYu}{Yu}

\end{document}